\documentclass[a4paper,fleqn]{cas-dc}

\usepackage[numbers]{natbib}

\def\tsc#1{\csdef{#1}{\textsc{\lowercase{#1}}\xspace}}
\tsc{WGM}
\tsc{QE}
\tsc{EP}
\tsc{PMS}
\tsc{BEC}
\tsc{DE}
\usepackage{microtype}
\RequirePackage[normalem]{ulem} 
\RequirePackage{color}\definecolor{RED}{rgb}{1,0,0}\definecolor{BLUE}{rgb}{0,0,1} 

\begin{document}
\let\WriteBookmarks\relax
\def\floatpagepagefraction{1}
\def\textpagefraction{.001}
\shorttitle{Microrheology of colloidal suspensions via Dynamic Monte Carlo simulations}
\shortauthors{F. A. Garc\'ia Daza et~al.}

\title [mode = title]{Microrheology of colloidal suspensions via Dynamic Monte Carlo simulations}                      



\author[1]{Fabi\'an A. {Garc\'ia Daza}}[orcid=0000-0002-8473-8349]
\cormark[1]
\ead{fabian.garciadaza@manchester.ac.uk}


\address[1]{Department of Chemical Engineering and Analytical Science, The University of Manchester, Manchester, M13 9PL, UK}

\author[2]{Antonio M. Puertas}[orcid=0000-0003-4127-1424]

\author[3]{Alejandro Cuetos}[orcid=0000-0003-2170-0535]

\address[2]{Department of Chemistry and Physics, University of Almer\'ia, 04120, Almer\'ia, Spain}

\address[3]{Department of Physical, Chemical and Natural Systems, Pablo de Olavide University, 41013, Sevilla, Spain}

\author[1]{Alessandro Patti}[orcid=0000-0002-7535-0000]
\cormark[1]
\ead{alessandro.patti@manchester.ac.uk}

\cortext[cor1]{Corresponding author}


\begin{abstract}
Understanding the rheology of colloidal suspensions is crucial in the formulation of a wide selection of industry-relevant products, such as paints, foods and inks. To characterise the viscoelastic behaviour of these soft materials, one can analyse the microscopic dynamics of colloidal tracers diffusing through the host fluid and generating local deformations and stresses. This technique, referred to as microrheology, links the bulk rheology of fluids to the microscopic dynamics at the particle scale. If tracers are subjected to external forces, rather than freely diffusing, it is called active microrheology. Motivated by the impact of microrheology in providing information on local structure in complex systems such as colloidal glasses, active matter or biological systems, we have extended the dynamic Monte Carlo (DMC) technique to investigate active microrheology in colloidal suspensions. The original DMC theoretical framework, able to accurately describe the Brownian dynamics of colloids at equilibrium, is here reconsidered and expanded to describe the effects of an external force pulling a tracer embedded in isotropic colloidal suspensions at different densities. To this end, we studied the dynamics of a spherical tracer dragged by a constant external force through a bath of spherical and rod-like particles of comparable size. We could extract valuable details on its effective friction coefficient, being constant at small and large values of the external force, but otherwise displaying a nonlinear behaviour that indicates the occurrence of a force-thinning regime. Our DMC simulation results are in excellent quantitative agreement with past Langevin dynamics simulations and theoretical works for the bath of spherical colloids. The bath of rod-like particles is studied in the isotropic phase, and displays an example where DMC is more convenient than Brownian or Langevin dynamics, in this case in dealing with particle rotation.
\end{abstract}




\maketitle

\section{Introduction}

Microrheology (MR) is a very intriguing technique that is applied to determine the local viscoelastic behaviour of a complex material by tracking the motion of small tracer particles (or probes) embedded in it \cite{MASON1995, CICUTA20071449, SQUIRES2010413}. Its range of applicability has been extended to the study of isotropic \cite{Habdas2004,MEYER2006,wilson2009,SRIRAM2010} and anisotropic \cite{Habibi2019, Paladugu20207556} colloidal suspensions, biopolymers \cite{Amblard19964470,Cribb20103311}, emulsions \cite{MEDRONHO2018225}, biological \cite{Feneberg2001284} and exotic \cite{Watts2014} systems. While in passive MR one can infer the viscoelastic behaviour of a soft material by following the free tracer's diffusion induced by the thermal fluctuations in the bath, in active MR a tracer is pulled by an external force. Consequently, passive MR captures the linear viscoelastic response of the material from the tracer motion induced by thermal fluctuations in the host system, and active MR addresses both the linear and nonlinear regimes by studying the response of the tracer to an external stimulus. In constant-force MR, the tracer velocity, $\langle v\rangle$, is calculated to estimate the effective friction of the bath, $\gamma_{\text{eff}}$, through the steady-state Stokes drag relationship $\textbf{F}_{\text{ext}}=\gamma_{\text{eff}}\langle v\rangle$, where $\textbf{F}_{\text{ext}}$ is the external force on the tracer, and the angular brackets refer to the steady-state ensemble average reached at long times. In this scenario, several theoretical works on a bath of spheres and equally large tracers, have predicted a plateau for the effective friction at small forces, corresponding to a linear relation between the applied force and the tracer velocity \cite{SQUIRES2005, Khair200673, HUANG2020293, GAZUZ2009, Gnann20111390}. This plateau is followed by a force-thinning, nonlinear regime, where friction decreases with increasing force, until a second plateau is reached at large forces. Although the microviscosity is the most direct observable in microrheology, additional relevant information of the bath can be obtained, such as its microscopic dynamics or heterogeneity 
\cite{Levine2001,Levine20001774}.

Molecular simulation has more recently emerged as a very powerful approach to study active MR of colloidal suspensions. Although Molecular Dynamics (MD) would perhaps be the optimal choice for the study of equilibrium and out-of-equilibrium dynamics of colloidal systems, the explicit inclusion of the solvent, along with the considerable length and time scales involved in the host fluid relaxation dynamics, would result in an excessively high computational cost. Mesoscale simulation techniques that employ an implicit solvent model are better suited to mitigate this challenge. In the pioneering Brownian dynamics (BD) simulations by \citeauthor{CARPEN2005} \cite{CARPEN2005}, the microscopic friction of monodisperse systems of hard spheres was calculated by monitoring the response of a spherical tracer dragged through the host particles at constant velocity or force. Similarly, Langevin dynamics simulations have been employed to investigate active MR in polydisperse systems of quasi-hard spheres forming concentrated phases with probes of the same \cite{GAZUZ2009, Puertas_2014} or different \cite{ORTS20198, ORTS2020052607} size as that of the bath particles. Also constant-force MR of a jammed suspension of soft particles has been determined by a simulation study \cite{MOHAN2014}. Furthermore, BD simulations have also been employed to gain insight into the mechanism of formation of clusters in nematic liquid crystals of rod-like particles incorporating a spherical tracer diffusing at constant velocity \cite{Wensink2006}. 

In this context, the dynamic Monte Carlo (DMC) simulation technique is an excellent alternative to BD to mimic the dynamics of colloidal suspensions. Its main advantage lies in its ability to capture the stochastic nature of colloids' dynamics with no need of integrating deterministic or stochastic equations of motion as in MD or BD simulations, respectively. By contrast,  DMC is fundamentally based on the Metropolis algorithm that is typically employed in standard Monte Carlo simulations. Although this would yield a non-deterministic progression of states, in the limit of small displacements and rotations (set according to the diffusion coefficients at infinite dilution), the DMC simulations can accurately replicate colloids' Brownian motion and thus result in time-dependent processes \cite{KIKUCHI1991, HEYES1998}. In recent years, different authors have applied DMC to the study of an ample spectrum of dynamical properties of colloidal systems. Sanz and co-workers studied the dynamics and nucleation in systems of quasi-hard and charged spheres as well as patchy particles, which have orientational degrees of freedom \cite{SANZ2010, Romano2011}. Similarly, \citeauthor{JabbariFarouji2012} performed DMC simulations to access the dynamics of hard spheres and thin platelets in dilute and concentrated regimes \cite{JabbariFarouji2012}. Almost simultaneously to these works, we proposed a DMC algorithm to study the dynamics of anisotropic particles in isotropic and liquid crystal phases \cite{PATTI2012}. In particular, we showed that, by rescaling an arbitrarily set MC time step with the acceptance rate of particle elementary moves, it is possible to recover a unique Brownian time scale that allows for a quantitative comparison with BD simulations. Over the last years, this DMC technique has been successfully applied and further expanded to investigate colloidal mixtures \cite{CUETOS2015, Chiappini2020PRL,TONTI2021116640}, systems undergoing transitory unsteady states \cite{CORBETT2018, Lebovka2019, Lebovka2019052135}, colloidal suspensions that exhibit phase coexistence \cite{GARCIA2020} and the equilibrium dynamics of especially exotic particles \cite{CUETOS2020, Chiappini2020}.

In this work, we reconsider and adapt the fundamentals of the DMC formalism to embrace the study of active MR of colloidal suspensions. More specifically, we consider the case of an external constant force acting on a tracer immersed in a bath of colloidal particles. To this end, two MC time steps are defined, one for the tracer and one for the bath particles. Any elementary move (tracer or bath particle) is linked to its corresponding MC time step through the Einstein relations and the associated diffusion coefficients at infinite dilution. We show that the MC time steps of tracer and host particles are not independent, but correlated \textit{via} their corresponding acceptance rates.  

This  paper is organised as follows: in Section \ref{sec2}, we describe the theoretical background of the DMC method and discuss the extension of the method to the case of an external force acting on a tracer. In section \ref{sec3}, we introduce the model and simulation methodology applied; in Section \ref{sec4}, we discuss the results obtained from DMC simulations of a tracer pulled by an external constant force through a colloidal bath. First a bath of spherical particles is considered, and the results are compared with Langevin dynamics simulations finding good agreement, and second, a bath of colloidal rods is studied. Finally, in Section \ref{sec5} we draw our conclusions.


\section{Theoretical Framework}
\label{sec2}

In this section, we discuss the fundamentals for the application of the DMC simulation technique to mimic the  dynamics of a colloidal suspension when a constant force is applied to a selected particle, here referred to as \textit{tracer}. At equilibrium, host (or bath) particles and tracer experience Brownian motion under the effect of the fluid thermal fluctuations. We stress that the fluid is not explicitly modelled and fluid-mediated hydrodynamic interactions are neglected. Our aim is to establish a consistent time scale linking the dynamical properties obtained \textit{via} the DMC method to the Brownian motion of the system in the presence of an external potential acting exclusively on the tracer. While the existence of a unique MC time scale that allows for a direct comparison between MC and BD simulations has been successfully established \cite{PATTI2012} and extended for an ample spectrum of systems and processes \cite{CUETOS2015,CORBETT2018, GARCIA2020,CUETOS2020}, it is still to clarify whether DMC can also be employed to investigate active MR in colloids. The methodology applied in this study is based on our former works on mono \cite{PATTI2012} and multicomponent \cite{CUETOS2015} colloidal suspensions. However, we here need to extend these original concepts to the case of an external force acting on a probe particle. In the following, we first present a concise overview of the theoretical background that is relevant to our discussion \cite{PATTI2012}, and then how this theory, and its extension to multicomponent systems \cite{CUETOS2015}, needs to be modified to the case when an external force is applied to a tracer.

\subsection{DMC Simulations for free Brownian motion}
\label{sec21}

In this section we provide the technical details of the DMC methodology for particles whose dynamics is governed by Brownian motion. First, we review the 1D case and then present the generalisation to an arbitrary number of degrees of freedom \cite{PATTI2012}. Let's consider a bath ($b$) particle, originally located at $x_b=0$, which is displaced to a new random position within the interval $\left[-\delta x,\delta x\right]$. According to the standard Metropolis-based MC algorithm, the acceptance probability of this move depends on the energy change, $\Delta U$, between new and old configurations of the system, and is determined by $\exp\left(-\beta\Delta U\right)$, where $\beta=1/k_{\text{B}}T$. This probability depends on the extent of the displacement, approaching unity as the displacement decreases. In the absence of external forces, the probability, $P_{\text{move},b}=\mathcal{A}_b/(2\delta x)$, of successfully moving a bath particle, with $ \mathcal{A}_{b}$ the acceptance rate, can be assumed to be constant over the interval $\left[-\delta x,\delta x\right]$. It should be noticed that there are specific cases where this assumption must be carefully evaluated, such as unsteady-state processes \cite{CORBETT2018} and space anisotropy \cite{GARCIA2020}. The particle mean displacement reads

\begin{equation}
    \label{eq1}
    \langle x_b\rangle =\int_{-\delta x}^{\delta x} x\:P_{\text{move},b} dx=0,
\end{equation}

\noindent which is null, as expected for Brownian motion of colloidal particles. By contrast, the mean square displacement (MSD) along a single MC step is given by,

\begin{equation}
    \label{eq2}
    \langle x_b^2\rangle =\int_{-\delta x}^{\delta x} x^2\:P_{\text{move},b} dx=\frac{\mathcal{A}_b\left(\delta x\right)^2}{3}.
\end{equation}

\noindent While in an MC step we attempt to update the position of a single particle, in an MC cycle we perform $N$ statistically independent MC steps, being $N$ the total number of particles in the system. In the case of $N$ particles performing $\mathcal{C}_{\text{MC}}$ cycles, the mean square displacement can be written as

\begin{equation}
    \label{eq3}
    \langle x_b^2\rangle  = \mathcal{C}_{\text{MC}} \frac{\mathcal{A}_b\left(\delta x\right)^2}{3}.
\end{equation}

\noindent This expression can be extended to higher dimensions. In a three-dimensional (3D) space the number of degrees of freedom increases to three in case of spherical particles, related to the displacement of their centre of mass, and to five for particles with axial symmetry, with two additional degrees of freedom associated to rotations. One can consider the most general case of $f$ degrees of freedom. A particle is displaced from the origin to a point $\xi=(\xi_1,\xi_2,\ldots,\xi_f)$ belonging to an $f$-dimensional hyperprism of sides $\left[-\delta\xi_k,\delta\xi_k\right]$, with $k=1,2,\ldots,f$. As in the previous case, the acceptance probability $\mathcal{A}_b$ is considered to be uniform over the hyperprism, and the corresponding normalised probability is $P_{\text{move},b}=\mathcal{A}_b/V_{\Xi}$, where $V_{\Xi}=\prod_i^f(2\delta\xi_k)$ is the volume of the hyperprism ${\Xi}$. Similarly to the 1D case, the mean displacement for each degree of freedom is found to be $\langle \xi_{b,k}\rangle  = 0$ for any $k\leq f$. By contrast, the MSDs of $N$ identical particles performing $\mathcal{C}_{\text{MC}}$ cycles take the form:

\begin{equation}
    \label{eq4}
    \langle \xi_{b,k}^2\rangle  = \mathcal{C}_{\text{MC}}\frac{\mathcal{A}_b\left(\delta \xi_k\right)^2}{3}.
\end{equation}

\noindent This result allows us to define an MC time scale, $t_{\text{MC},b}$, that can be directly related to the time scale of a BD simulation, $t_{\text{BD}}$. To this end, we define an MC step through the relation 

\begin{equation}
 \left(\delta \xi_k\right)^2 = 2D_k\delta t_{\text{MC,b}},
 \label{eq5}
\end{equation}

\noindent where $D_k$ is the diffusion coefficient at infinite dilution of the $k$th degree of freedom, and $\delta t_{\text{MC,b}}$ is the time needed to perform an MC cycle. The Einstein relation

\begin{equation}
    \label{eq6}
    \langle \xi_{b,k}^2\rangle =2D_k t_{\text{BD}}
\end{equation}

\noindent allows to establish a link with the BD time scale by combining the expressions for $\langle \xi_{b,k}^2\rangle $ in Eqs.~\ref{eq4} and \ref{eq6} with the MC relation in Eq.~\ref{eq5} for $(\delta \xi_k)^2$:

\begin{equation}
    \label{eq7}
    t_{\text{BD}} = \mathcal{C}_{\text{MC}} \delta t_{\text{BD}}= \frac{\mathcal{A}_b}{3}\mathcal{C}_{\text{MC}}\delta t_{\text{MC},b},
\end{equation}

\noindent where $\delta t_{\text{BD}}$ is the elementary time step in the BD timescale. This equation offers a relation between the time scale of DMC and BD simulations. More specifically, any arbitrary value set for $\delta t_{\text{MC},b}$ can then be rescaled with  $\mathcal{A}_b$  to recover the BD time scale. Therefore, Eq.~\ref{eq7} provides a unique time scale for DMC simulations.   

\subsection{DMC simulations of colloids in a 1D potential}
\label{sec22}

When a system is perturbed by an external force, the acceptance rate is no longer independent from the extent of displacement nor uniform in the hyperprism of sides $\left[-\delta \xi_k, \delta \xi_k\right]$. This will affect the average quantities calculated by the integrals in Eqs.\,\ref{eq1} and \ref{eq2} and modified expressions of $\langle  x\rangle $ and $\langle  x^2\rangle $ are expected. For the sake of clarity, the trivial case of a particle moving in one dimension subject to an external potential is discussed first.  Let us consider a tracer (\textit{t}) located at a point $x$ and an external force $\mathbf{F}_{\text{ext}}=F_{\text{ext}}\hat{\textbf{x}}=-\nabla U_{\text{ext}}(x)$ acting on it. Similarly to previous works \cite{KIKUCHI1991, HEYES1998, SANZ2010}, the particle is allowed to displace to a new position within the interval $\left[-\delta x,\delta x\right]$ provoking a change in the potential energy $\Delta U_{\text{ext}}(x)$. If the displacement is in the same direction of the force, $\Delta U_{\text{ext}}(x)<0$, the move is accepted with probability $P_{\text{move},t}=\mathcal{A}_0/(2\delta x)$,  and the particle is allowed to occupy its new position. $\mathcal{A}_0$ is the probability of accepting the displacement of the tracer considering its interaction with the neighboring particles. By contrast, for attempted moves in the opposite direction to the force, then $\Delta U_{\text{ext}}(x)>0$ and the move is accepted with probability $P_{\text{move},t}=\mathcal{A}_0\:e^{-\beta\Delta U_{\text{ext}}(x)}/(2\delta x)$. We are therefore assuming that the external force favours the displacement of the particle towards the positive direction of the interval.
In the present work, we consider the potential to exhibit a smooth behaviour on the spatial coordinate $\hat{x}$, i.e. $\beta\Delta U_{\text{ext}}(x) = -\beta F_{\text{ext}}x$. It follows that any average quantity can be derived as an expansion in power series of $x$. In particular, the mean displacement of the tracer is given by

\begin{align}
\label{eq8}
    \langle  x_t\rangle  &= \frac{\mathcal{A}_0}{2\delta x}\int_{-\delta x}^{0} xe^{\beta F_{\text{ext}} x}dx + \frac{\mathcal{A}_0}{2\delta x}\int_{0}^{\delta x} x\:dx\nonumber\\
        & \simeq \frac{\beta F_{\text{ext}} \mathcal{A}_0\left(\delta x\right)^2}{6}\left(1-\frac{3}{8}\beta F_{\text{ext}}\delta x\right).
\end{align}

\noindent In the above equation we have neglected the terms $\mathcal{O}((\beta F_{\text{ext}}\delta x)^2)$ that stem from the expansion of the exponential. This assumption is valid as long as the following condition is fulfilled: 

\begin{equation}
    \beta F_{\text{ext}} \delta x\ll1.
    \label{eq9}
\end{equation}

\noindent Extending the result in Eq.~\ref{eq8} to the case of the particle performing $\mathcal{C}_{\text{MC}}$ cycles, the mean displacement reads

\begin{equation}
\label{eq10}
    \langle  x_t\rangle  \simeq \mathcal{C}_{\text{MC}}\frac{\beta F_{\text{ext}} \mathcal{A}_0\left(\delta x\right)^2}{6}\left(1-\frac{3}{8}\beta F_{\text{ext}}\delta x\right).
\end{equation}

\noindent This result differs from the case of systems in absence of external forces \cite{PATTI2012, CUETOS2015} (see Eq.\,\ref{eq1}), where particles move by pure Brownian motion. Similarly, the expansion of the MSD is found to be

\begin{align}
\label{eq11}
    \langle  x_t^2\rangle  &= \frac{\mathcal{A}_0}{2\delta x}\int_{-\delta x}^{0} x^2e^{\beta F_{\text{ext}} x}dx + \frac{\mathcal{A}_0}{2\delta x}\int_{0}^{\delta x} x^2 dx\nonumber\\
    & \simeq \frac{\mathcal{A}_0\left(\delta x\right)^2}{3} \left(1-\frac{3}{8}\beta F_{\text{ext}} \delta x \right).
\end{align}

\noindent Again, this result can be extended to the case of a particle performing $\mathcal{C}_{\text{MC}}$ cycles: 

\begin{equation}
    \label{eq12}
    \langle  x_t^2\rangle \simeq \mathcal{C}_{\text{MC}}\frac{\mathcal{A}_0\left(\delta x\right)^2}{3} \left(1-\frac{3}{8}\beta F_{\text{ext}} \delta x \right).
\end{equation}

\noindent We note that the presence of the external force results in additional terms in the MSD as compared to Eq.\,\ref{eq3}. On the other hand, we observe that the relative corrections in the mean and mean square displacements are the same. As expected, the expressions given in Eqs.\,\ref{eq1} and \ref{eq3} are recovered as $F_{\text{ext}}\rightarrow0$. Expanding the probability to accept the tracer move, $\mathcal{A}_t$, leads to

\begin{align}
\label{eq13}
    \mathcal{A}_t &= \frac{\mathcal{A}_0}{2\delta x}\int_{-\delta x}^{0}e^{\beta F_{\text{ext}} x}dx + \frac{\mathcal{A}_0}{2\delta x}\int_{0}^{\delta x}dx\nonumber\\
    & \simeq \mathcal{A}_0\left(1-\frac{\beta F_{\text{ext}} \delta x}{4} \right).
\end{align}

\noindent As can be noticed, the acceptance probability does not have the same expansion in powers of $\delta x$ as in the case of $\langle  x_t\rangle $ and $\langle  x_t^2\rangle $. This was also observed in the work by \citeauthor{SANZ2010} \cite{SANZ2010}, and more recently by \citeauthor{CORBETT2018} \cite{CORBETT2018}. By combining Eqs.~\ref{eq10}, \ref{eq12} and \ref{eq13}, one obtains

\begin{align}
\label{eq14}
    \langle  x_t\rangle  &= \mathcal{C}_{\text{MC}}\frac{\beta F_{\text{ext}} \left(\delta x\right)^2}{6}\left(\frac{3}{2}\mathcal{A}_t-\frac{1}{2}\mathcal{A}_0\right),\\    
\label{eq15}
    \langle  x_t^2\rangle  &= \mathcal{C}_{\text{MC}}\frac{\left(\delta x\right)^2}{3}\left(\frac{3}{2}\mathcal{A}_t-\frac{1}{2}\mathcal{A}_0\right).
\end{align}

\noindent Again, the presence of the external force affects the acceptance rate leading to a different rescaling represented by the terms involving $\mathcal{A}_t$ in both $\langle  x_t\rangle $ and $\langle x_t^2\rangle$. It should be noted that, for small displacements of the tracer, $\mathcal{A}_0\rightarrow 1$ and one thus obtains $3\mathcal{A}_t/2-1/2\approx\mathcal{A}_t$ and, consequently, $\langle x_t^2\rangle\approx \mathcal{C}_{\text{MC}} \mathcal{A}_t \delta x^2/3$, being the usual expression found in different DMC studies of purely Brownian systems \cite{PATTI2012, CUETOS2020, CUETOS2015}. Similarly, \citeauthor{SANZ2010} \cite{SANZ2010} showed that corrections on the acceptance rate tend to vanish as the dimensionality of the force acting on the particles increases. This was subsequently corroborated by \citeauthor{CORBETT2018} \cite{CORBETT2018} when studying systems of rod-like particles undergoing a field-induced phase transition.

Eqs.\,\ref{eq14} and \ref{eq15} allow us to define a time scale in DMC simulations, $t_{\text{MC},t}$, that can be related to the time scale in BD simulations, $t_{\text{BD}}$. As discussed in Section \ref{sec21}, a common and efficient strategy to relate the displacements and rotations of a particle to a consistent temporal scale is through the Einstein relation, which is equivalent to the Langevin equation at long times \cite{einstein1956investigations}. If the MC moves are assumed to be statistically independent, time and space can be related through the diffusion coefficients at infinite dilution by the equation $\left(\delta x\right)^2\simeq 2D_t\delta t_{\text{MC},t}$, where $D_t$ is the diffusion coefficient of the tracer particle, and $\delta t_{\text{MC},t}$ is the time a particle takes to perform an MC cycle in the MC timescale. By combining this result with Eq.\,\ref{eq14}, the following result is obtained:

\begin{equation}
\label{eq16}
    \langle x_t\rangle = \frac{1}{3}D_t\mathcal{C}_{\text{MC}}\beta F_{\text{ext}} \left(\frac{3}{2}\mathcal{A}_t-\frac{1}{2}\mathcal{A}_0\right)\delta t_{\text{MC},t}.    
\end{equation}

\noindent Since the Langevin equation of a particle in 1D yields 

\begin{equation}
\label{eq17}
    \langle x_t\rangle  = D_t\beta F_{\text{ext}} t_{\text{BD}},
\end{equation}

\noindent by equating Eqs.\,\ref{eq16} and \ref{eq17}, a relation between MC and BD time scales is obtained:

\begin{equation}
    \label{eq18}
    t_\text{BD} = \mathcal{C}_{\text{MC}}\delta t_\text{BD} = \frac{1}{3}\left(\frac{3}{2}\mathcal{A}_t-\frac{1}{2}\mathcal{A}_0\right)\mathcal{C}_{\text{MC}}\delta t_{\text{MC},t}.
\end{equation}

\noindent This equation provides a direct link between the time scales in DMC and BD simulations in the presence of external potentials. The importance of this result, together with the relation in Eq.~\ref{eq7}, is that in a system containing a tracer (or tracers) pulled by an external force and bath particles, whose trajectory is only determined by thermal motion, the BD time step is the same for the two populations, but the MC time step is population-dependent. In this study we assume that the tracer displacements are sufficiently small such that $\mathcal{A}_0 \rightarrow 1$. Combining these results we obtain the following relationship:

\begin{equation}
    \label{eq19}
    \left(\frac{3}{2}\mathcal{A}_t-\frac{1}{2}\right)\delta t_{\text{MC},t} = \mathcal{A}_b\delta t_{\text{MC},b}
\end{equation}

\noindent We made sure that this assumption was indeed valid by explicitly estimating $\mathcal{A}_0$ in systems with the smallest acceptance rates, being approximately 0.8, and observed that the results remained unaffected. The question that arises now is how large the elementary MC time steps could be. While the bath particles can access a broad range of time steps as previously stressed in Ref.~\cite{PATTI2012}, the presence of an external force on the probe particle influences the choice of its MC time step. In the 1D case, any average quantity is expressed as an expansion in terms of the external force and the maximum displacement as in Eqs.\,\ref{eq10} and \ref{eq12}. To ensure that high-order terms in $\mathcal{O}(\beta F_{\text{ext}}\delta x)$ do not invalidate the approximations in Eqs.~\ref{eq8}, \ref{eq10}-\ref{eq13}, it is strictly necessary to fulfil the condition in Eq.~\ref{eq9}. Consequently, the MC time step of the tracer which depends on the maximum displacement ($\left(\delta x\right)^2\simeq2D_t\delta t_{\text{MC},t}$) is force dependent as $\delta t_{\text{MC},t}\ll (\beta F_{\text{ext}}\sqrt{2D_t})^{-1}$. This suggests that for small forces the MC time step of the tracer can be drawn from a wide range of values. By contrast, the magnitude of the time step reduces as the force increases.   


\section{Model and Methodology}
\label{sec3}

We have studied constant-force MR of two colloidal suspensions comprising $N$ bath particles and 1 tracer. The former are either quasi-hard spherical particles with diameter $\sigma$ or hard spherocylinders (HSCs) with length-to-diameter ratio $L^*\equiv L/\sigma=5$. In both cases, the tracer is a spherical particle of size $\sigma$. Active MR of fluids of quasi-hard spheres has already been studied by Langevin dynamics simulation \cite{Puertas_2014} and is here taken as a reference to test the validity of our DMC method. In particular, quasi-hard spheres interact via the following potential

\begin{equation}
 \label{eq20}
 U_{ij}=k_{\text{B}}T\left(\frac{r_{ij}}{\sigma} \right)^{-36},
\end{equation}

\noindent where $r_{ij}$ is the centre-to-centre distance between spheres $i$ and $j$ (bath or tracer). This quasi-hard-core potential closely resembles that of hard spheres, but allows integration of Newton equations in Langevin dynamics simulation. This limitation is not relevant in DMC simulations, where time trajectories are generated by purely stochastic events. As such, we decided to model the rod-like particles of our second set of systems as HSCs, which interact via the following hard-core (excluded-volume) potential 

\begin{equation}
  \label{eq21}
  U_{ij}=
  \begin{cases}
    0,& \text{if } d_{m}(\textbf{r}_{ij},\textbf{u}_{i},\textbf{u}_{j})\ge\sigma\\
    \infty,& \text{if } d_{m}(\textbf{r}_{ij},\textbf{u}_{i},\textbf{u}_{j}) < \sigma,
  \end{cases}
\end{equation}

\noindent where $U_{ij}$ is the interaction potential between two generic particles of species $i$ and $j$ (rod-rod or rod-tracer pairs) with orientations $\textbf{u}_{i}$ and $\textbf{u}_{j}$, respectively, and whose geometric centers are separated by the vector $\textbf{r}_{ij}$. The minimum distance, $d_m$, between particles $i$ and $j$ is calculated following Ref.\,\cite{VEGA199455}. Energy, length and time are, respectively, given in units of $k_{\text{B}}T$, $\sigma$ and $\tau=\sigma^2/D_0$, with $D_0=k_{\text{B}}T/(\eta\sigma)$ a diffusion constant and $\eta$ the viscosity coefficient of the solvent. In addition to the interaction established with the particles of the bath, the tracer is pulled by an external force $\mathbf{F}_{\text{ext}}$ and dragged through the simulation box. We assume this force to be constant and parallel to the \textit{x}-axis. In particular,
\begin{equation}
    \label{eq22}
    \mathbf{F}_{\text{ext}}=Pe\frac{k_{\text{B}}T}{a}\hat{\textbf{x}},
\end{equation}
where $a=\sigma/2$ is the tracer radius and $Pe$ is the P\'eclet number, which quantifies the relevance of the external force with respect to thermal forces. 

\begin{figure}
	\centering
	\includegraphics[scale=1]{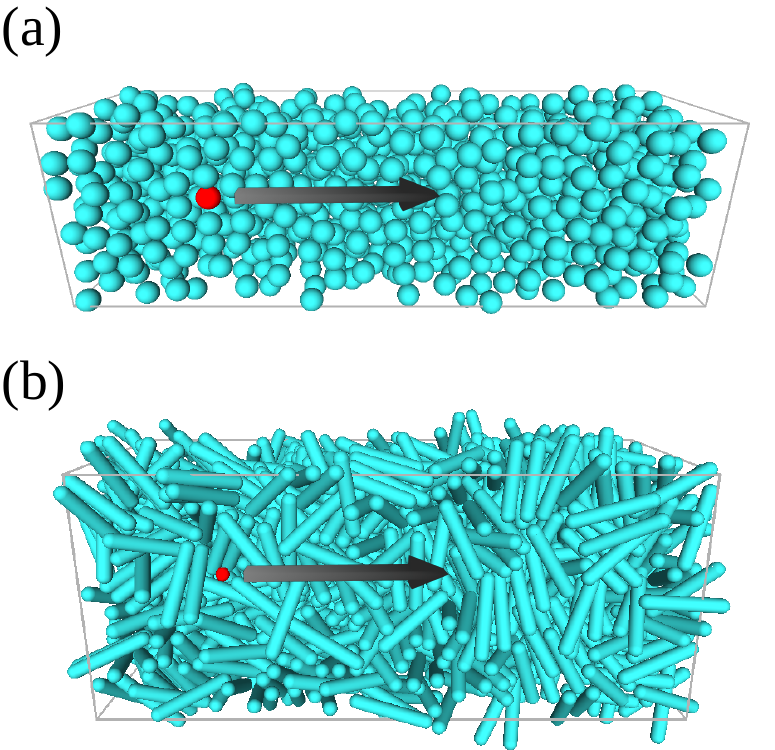}
	\caption{(colour on-line) Typical configurations of a colloidal suspension of (a) quasi-hard spheres and (b) hard spherocylinders. The arrow indicates the direction of the external force applied to the spherical tracer (in red). Both systems are at the same temperature. The volume fraction of the baths are $\phi=0.30$ and $\phi=0.38$ for spheres and rods, respectively.}
	\label{FIG:1}
\end{figure}

The systems studied have been equilibrated by performing standard MC simulations in the \textit{NVT} ensemble at $k_{\text{B}}T=1.0$. Simulations were performed in elongated simulation boxes of sizes $L_x$ and $L_y=L_z$ with periodic boundaries. For the bath of spheres, $L_y=L_z=8\sigma$, and $L_x$ varies between $5.1 L_y$ and $2.0 L_y$ for volume fractions $\phi=\frac{4}{3}\pi a^3n_c$ between $0.20$ and $0.50$, respectively, being $n_c$ the bath particles number density. Similarly, for the bath of rods, $L_y=L_z=17\sigma$, and $L_x$ ranges from $4.5 L_y$ to $2.4 L_y$ for $\phi=\big(\frac{4}{3}\pi a^3+\pi a^2L\big)n_c$ between $0.20$ and $0.38$, respectively. All systems consisted of 1000 bath particles and 1 tracer. Equilibrium was considered achieved when the total energy reached a steady value within reasonable statistical fluctuations. We then selected one of the equilibrated configurations as the initial one for the DMC production runs. Exemplary snapshots of systems of spheres and rods incorporating a spherical tracer are provided in Fig.~\ref{FIG:1}.      

\subsection{DMC Simulations}

\sloppy In the production stage, we performed DMC simulations in the $NVT$ ensemble in elongated boxes with periodic boundary conditions. Since our aim is mimicking Brownian motion, unphysical moves usually employed in standard MC simulations, such as swaps, jumps or cluster moves, are not allowed here. Simultaneous displacements and rotations (the latter attempted on rods only) of randomly selected particles, were accepted or rejected according to the standard Metropolis algorithm, that is with probability $\min\left[1,\exp\left(-\Delta U/k_{\text{B}}T\right)\right]$, where $\Delta U$ is the change in energy resulting from the proposed move. Shall the move be accepted, the energy of the system is increased by $\Delta U$. The displacement of the spheres in the bath is decoupled into three terms: $\delta\textbf{r}_s = X_x\hat{\textbf{x}} + X_y\hat{\textbf{y}} + X_z\hat{\textbf{z}}$, where $X_l$, with $l=x,y,z$, represents random numbers satisfying the condition $|X_l|\leq\delta r_s^{\text{max}}$, being $\delta r_s^{\text{max}}$ the maximum displacement of a spherical particle along direction $l$. Similarly, the displacement of the rods is $\delta\textbf{r}_r = X_\parallel\hat{\textbf{u}}_r + X_{\perp,1}\hat{\textbf{v}}_{r,1} + X_{\perp,2}\hat{\textbf{v}}_{r,2}$, where $\hat{\textbf{u}}_r$ is a unit vector parallel to the rod orientation and $\hat{\textbf{v}}_{r,m}$, with $m=1,2$, are two randomly chosen unit vectors that are perpendicular to each other and to $\hat{\textbf{u}}_r$. The magnitude of the displacements is chosen at random from uniform distributions fulfilling $|X_\parallel|\leq\delta r_\parallel^{\text{max}}$ and $|X_{\perp,m}|\leq\delta r_\perp^{\text{max}}$, where $\delta r_\parallel^{\text{max}}$ and $\delta r_\perp^{\text{max}}$ are the maximum elementary displacements parallel and perpendicular to $\hat{\textbf{u}}_r$, respectively. For the rotations, the orientation vector changes from $\hat{\textbf{u}}_r$ to $\hat{\textbf{u}}_r + \delta\hat{\textbf{u}}_r$ with $\delta\hat{\textbf{u}}_r=Y_{\varphi,1}\hat{\mathbf{w}}_{r,1} + Y_{\varphi,2}\hat{\mathbf{w}}_{r,2}$. Vectors $\hat{\mathbf{w}}_{r,m}$ are randomly chosen and are perpendicular to each other and to $\hat{\textbf{u}}_r$. The random numbers $Y_{\varphi,m}$ are chosen to satisfy the condition $|Y_{\varphi,m}|\leq\delta\varphi^{\text{max}}$, where $\delta\varphi^{\text{max}}$ is the maximum rotation of the major axis of the rods. Finally, the displacement of the spherical tracer reads $\delta\textbf{r}_t = X_{\parallel}^{t}\hat{\textbf{u}}_t + X_{\perp,1}^{t}\hat{\textbf{v}}_{t,1} + X_{\perp,2}^{t}\hat{\textbf{v}}_{t,2}$, where $\hat{\textbf{u}}_t$ is a vector parallel to the external force $\mathbf{F}_{\text{ext}}$, whereas $\hat{\textbf{v}}_{t,m}$ are two randomly chosen vectors that are perpendicular to each other and to $\hat{\textbf{u}}_t$. The magnitude of the displacement of the tracer is chosen at random with the conditions $|X_{\parallel}^{t}|\leq\delta r_{\parallel}^{\text{max},t}$ and $|X_{\perp,m}^{t}|\leq \delta r_{\perp}^{\text{max},t}$, where $\delta r_{\parallel}^{\text{max},t}$ and $\delta r_{\perp}^{\text{max},t}$ are the maximum displacements parallel and perpendicular to $\hat{\textbf{u}}_t$, respectively. Maximum displacements and rotations of the particles in the bath can be linked to their diffusion coefficients at infinite dilution via the following Einstein relations:

\begin{align}
    \label{eq23}
    \delta r_s^{\text{max}} &= \sqrt{2D_s\delta t_{\text{MC},s}}\\
    \label{eq24}
    \delta r_{\perp}^{\text{max}} &= \sqrt{2D_{r,\perp}\delta t_{\text{MC},r}}\\
    \label{eq25}
    \delta r_{\parallel}^{\text{max}} &= \sqrt{2D_{r,\parallel}\delta t_{\text{MC},r}}\\
    \label{eq26}
    \delta \varphi^{\text{max}} &= \sqrt{2D_{r,\varphi}\delta t_{\text{MC},r}}
\end{align}

\noindent While the maximum displacement of the tracer in the direction perpendicular to the external force is governed by Brownian motion, that parallel to this force must include an inertial term. More specifically:

\begin{align}
    \label{eq27}
    \delta r_{\perp}^{\text{max},t} &= \sqrt{2D_t\delta t_{\text{MC},t}}\\
    \label{eq28}
    \delta r_{\parallel}^{\text{max},t} &= \sqrt{2D_{t}\delta t_{\text{MC},t}+\left(D_t\beta F_{\text{ext}} \delta t_{\text{MC},t}\right)^2}
\end{align}

\noindent The diffusion coefficient at infinite dilution of bath and tracer spherical particles is obtained from the Stokes-Einstein equation. In this work we have considered the same size for the bath and tracer particles. Consequently, their diffusion coefficients are related by:

\begin{equation}
    \label{eq29}
    \frac{D_t}{D_0} = \frac{D_s}{D_0}=\frac{1}{3\pi}.
\end{equation}

\noindent To estimate the rod translational and rotational diffusion coefficients at infinite dilution, we have applied the analytic results based on the induced forces method by \citeauthor{BONETAVALOS1994193} \cite{BONETAVALOS1994193}. In particular:

\begin{align}
    \label{eq30}
    \frac{D_{r,\perp}}{D_0}&=\frac{\ln(2/\epsilon)-1/2-I^{tt}}{2\pi/\epsilon}\\
    \label{eq31}
    \frac{D_{r,\parallel}}{D_0}&=\frac{\ln(2/\epsilon)-3/2-I^{tt}}{\pi/\epsilon}\\
    \label{eq32}
    \frac{D_{r,\varphi}}{D_0}&=3\frac{\ln(2/\epsilon)-11/6-I^{rr}}{\pi\sigma^2/\left(2\epsilon\right)^3}
\end{align}

\noindent where $1/\epsilon=2\left(L^*+1\right)$, $I^{tt}\equiv\frac{1}{2}\int_{-1}^{1}dx \ln h(x)$ and $I^{rr}\equiv\frac{3}{2}\int_{-1}^{1}dx x^2\ln h(x)$, with $h(x)=(1-2x^{2n})^{1/2n}$ a parametric function used to model particles with symmetry of revolution. By choosing $n=8$, the resulting particle shape approximates very well that of a spherocylinder. Thereby $I^{tt}\simeq-0.0061$ and $I^{rr}\simeq-0.017$. The complete set of the diffusion coefficients at infinite dilution used in Eqs.\,\ref{eq29}-\ref{eq32} are given in Table~\ref{tbl1}.  

\begin{table}[width=.9\linewidth,cols=4,pos=h]
\caption{Diffusion coefficients at infinite dilution of spheres with diameter $\sigma$, and spherocylinders with aspect ratio $L^*=5$, calculated from Eqs.\,\ref{eq29}-\ref{eq32}.} \label{tbl1}
\begin{tabular*}{\tblwidth}{@{} LLLL@{} }
\toprule
$D_{t,s}/D_0$ & $D_{r,\perp}/D_0$ & $D_{r,\parallel}/D_0$ & $D_{r,\varphi}/D_0$\\
\midrule
$1.061\times10^{-1}$ & $3.560\times10^{-2}$ & $4.467\times10^{-2}$ & $6.020\times10^{-3}$\\
\bottomrule
\end{tabular*}
\end{table}

\begin{figure}[h!]
	\centering
		\includegraphics[scale=0.95]{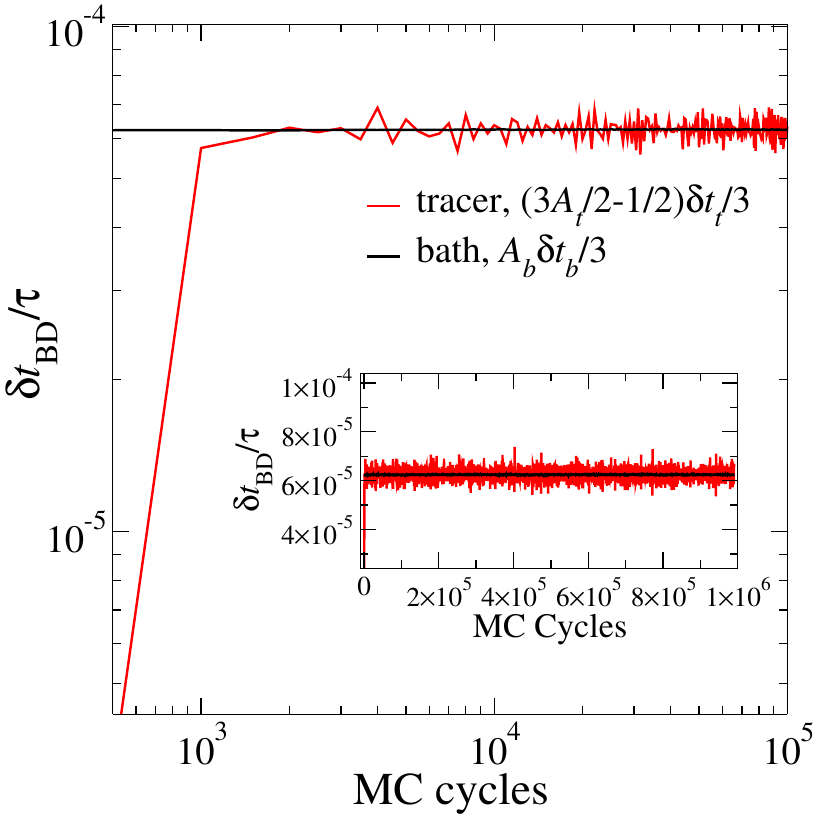}
	\caption{(colour on-line) Convergence of the rescaled MC time step of the tracer $\left(\rm 3\mathcal{A}_t/2 - 1/2\right)\delta t_{\text{MC,t}}/3$ (red solid line) to $\rm \mathcal{A}_b\delta t_{\text{MC,b}}/3$ (black solid line) according to Eq.\,\ref{eq19}. The two rescaled time steps converge to a steady value which corresponds to the elementary BD time step. The system consists of a bath of spheres with volume fraction $\phi=0.50$ and an external force ($Pe=10$) pulling the tracer.}
	\label{FIG:2}
\end{figure}

The maximum displacements and rotations are set according to the elementary MC time steps: $\delta t_{\text{MC},b}$ with $b=s,r$, for bath particles and $\delta t_{\text{MC},t}$ for the tracer. As established by Eq.\,\ref{eq19}, bath and tracer time scales are related by their corresponding acceptance rates. One can therefore arbitrarily set a fixed value of the time step of either the bath particles or tracer, and obtain the other by convergence of Eq.\,\ref{eq19} by running a preliminary DMC simulation. We have chosen to keep constant the time step of the bath, $\delta t_{\text{MC},b}$, within the range $10^{-7}\tau - 10^{-1}\tau$, depending on the volume fraction and force intensity. The tracer time step, $\delta t_{\text{MC},t}$, has been calculated by updating the instantaneous values of $\rm \mathcal{A}_b$ and $\rm  \mathcal{A}_t$ in Eq.\,\ref{eq19}. In particular, we initially set $\delta t_{\text{MC},b}>\delta t_{\text{MC},t}$ and recalculated $\delta t_{\text{MC},t}$ every 500 MC cycles until $\delta t_{\text{MC,t}} = 2\mathcal{A}_b \delta t_{\text{MC},b}/\left(3\mathcal{A}_t - 1\right)$. A visual convergence of Eq.\,\ref{eq19} at $Pe=10$ and $\phi=0.50$ is presented in Fig.\,\ref{FIG:2}, where the scaled tracer time step $\left(\rm 3\mathcal{A}_t/2 - 1/2\right)\delta t_{\text{MC,t}}$ equals $\rm \mathcal{A}_b\delta t_{\text{MC,b}}$ in approximately 2000 MC cycles. We stress that this procedure is essential to establish a unique BD time scale for all particles and consistently compare their dynamics on the basis of this time scale. Following this preliminary computation, the MC time steps as well as the maximum displacements and rotations are kept constant during the DMC production runs.

To quantitatively validate our results, we have calculated the effective friction coefficient at different $Pe$ numbers and bath volume fractions and then compared it with the values obtained, under identical conditions, by recent Langevin dynamics simulation \cite{GAZUZ2009, Puertas_2014}. To estimate the friction coefficient, we made use of the Stokes drag relation, which reads

\begin{equation}
\label{eq33}
    \frac{\gamma_\text{eff}}{\gamma_0}=\frac{F_\text{ext}}{6\pi\eta a\langle v\rangle },
\end{equation}

\noindent where $\gamma_0=6\pi\eta a$ is the friction coefficient of the medium, and $\langle v\rangle $ the long-time mean velocity of the tracer. At the beginning of each time trajectory, a random particle from the bath is assigned the status of tracer and the force $(F_{\text{ext}},0,0)=(Pek_{\text{B}}T/a,0,0)$ is applied on it. In case of a bath of rod-like particles, after randomly choosing a spherocylinder and converting it into the spherical tracer, a $10^5$-cycle MC equilibration is carried out to obtain uncorrelated trajectories. Once the equilibration is complete, the external force is exerted on the probe particle and its trajectory is monitored. When the tracer has moved a distance larger than $3L_x/4$, another selection is performed and a new trajectory starts. The long-time steady velocity of the tracer, $\langle v\rangle $, is calculated as the slope of the averaged tracer displacement with respect to $t_{\text{BD}}$. Sampling was performed over 400 independent time trajectories for each value of the external force. On average, at low $Pe$ we performed between $1\times10^4$ and $2\times10^6$ MC cycles for each tracer trajectory for the lowest and highest volume fractions, respectively. By contrast, at high $Pe$ the total number of MC cycles per trajectory was set to $6\times10^6$ for all the volume fractions under study.


\section{Results}
\label{sec4}

This section has a twofold aim: (\textit{i}) providing a sound evidence that our DMC simulation technique can be employed to investigate the nonlinear viscoelastic response of colloidal suspensions, and (\textit{ii}) applying DMC to study the active nonlinear MR of a suspension of colloidal rods. To this end, we first validate the DMC theoretical formalism by bench-marking our simulation results against existing Langevin dynamics simulation of quasi-hard-sphere particles \cite{GAZUZ2009,Puertas_2014}. Then, we report on the motion of a spherical tracer under constant-force steady driving in an isotropic bath of hard spherocylinders.


\subsection{Bath of colloidal quasi-hard spheres}

\begin{figure}[h!]
	\centering
		\includegraphics[scale=0.95]{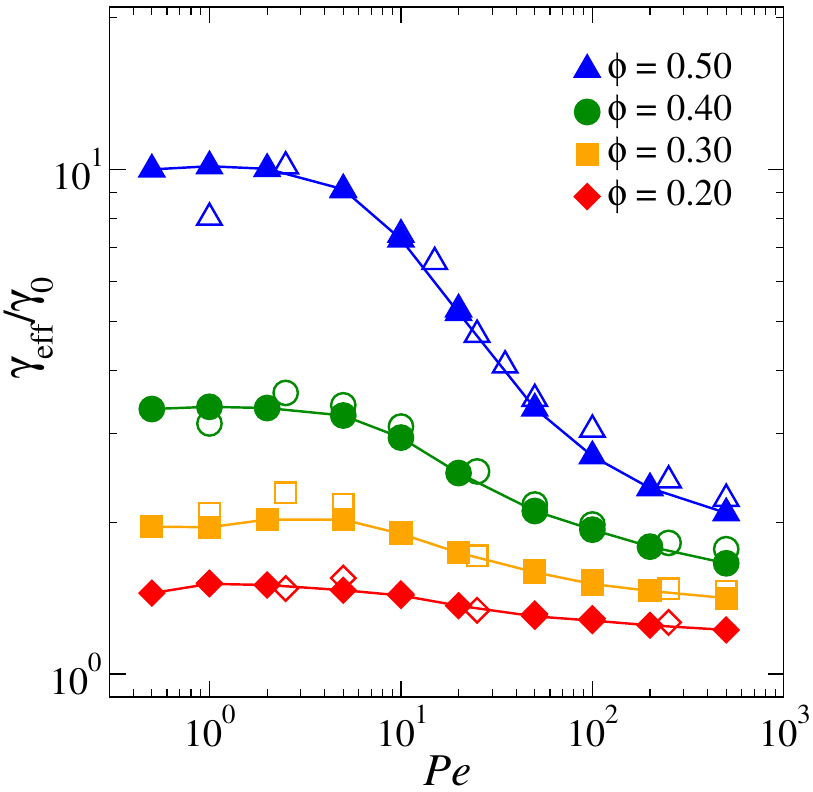}
	\caption{(colour on-line) Effective friction coefficient of quasi-hard spheres of the same size as the probe particle. Solid symbols and lines indicate DMC simulations results, while empty symbols refer to Langevin dynamics simulation results from Refs.~\cite{GAZUZ2009,Puertas_2014}. Solid lines are guides for the eye.}
	\label{FIG:3}
\end{figure}

To analyse the impact of density and force intensity on the effective friction coefficient, we have considered a wide spectrum of bath volume fractions and P\'eclet numbers. In particular, $\phi=0.20,\:0.30,\:0.40$, and $0.50$ and $0.5\leq Pe\leq 500$. Across the complete set of volume fractions, quasi-hard spheres are in the isotropic phase, being the glass transition at $\phi_c\approx0.595$ \cite{Voigtmann2004061506}. Details of the systems studied in this section are shown in Table S1 of the Supplementary Material. The effective friction coefficient, $\gamma_{\text{eff}}$, of a system of quasi-hard spheres against the force-based Péclet number is shown in Fig.~\ref{FIG:3}, where we also report the results of Langevin dynamics simulations from the literature \cite{GAZUZ2009, Puertas_2014}. 
The agreement between the two simulation techniques is excellent across the three regimes that we now discuss in detail. At low $Pe$, the presence of the tracer has a very mild, almost negligible effect on the equilibrium configuration of the bath system. This results in a linear relation between the external force exerted on the tracer and its velocity. Consequently, within such a linear regime, a constant value of the friction coefficient is observed. Upon increasing force intensity, that is at  intermediate $Pe$, the microstructure of the transient cage surrounding the tracer is gradually distorted, leading to \textit{force thinning}, where the effective friction coefficient decreases as $Pe$ is increased. Such a force thinning effect, analogous to the decrease of shear viscosity with increasing shear rate (i.e. shear thinning), has been predicted by theory \cite{SQUIRES2005} and simulations \cite{GAZUZ2009,CARPEN2005,Puertas_2014} and observed experimentally \cite{MEYER2006, wilson2009, SRIRAM2010}. In agreement with these former studies, we also observe a more pronounced force thinning at significantly large volume fractions (\textit{e.g.} $\phi=0.50$), where the limited particle mobility results into an especially large effective friction coefficient.
Upon further increasing the force intensity, say at $Pe>10^2$ depending on $\phi$, a second linear-response regime is found. Although Fig.~\ref{FIG:3} does not display a fully developed plateau at large $Pe$, it is evident, especially at $\phi \le 0.30$, that $d\gamma_{\text{eff}}/dPe$ is gradually approaching zero. Nevertheless, to the extent that the condition in Eq.\,(9) is fulfilled, simulations at external forces sufficiently large to observe a linear regime require remarkably small timesteps since $\delta t_{{\text{MC}},t}\ll (\beta F_{\text{ext}}\sqrt{2D_t})^{-1}$. According to our calculations (see Table S1 of the Supplementary Material) above $Pe=500$ the optimal timesteps are expected to be lower than $10^{-7}\tau$. The occurrence of a plateau at large $Pe$ is consistent with the established theoretical framework proposed by \citeauthor{SQUIRES2005} \cite{SQUIRES2005}. In this limit, advection is dominant over thermal motion, except in a thin layer surrounding the tracer, whose thickness is proportional to $\sigma Pe^{-1}$. This results into an entropic reactive force of the bath of the order $\sim Pe k_{\text{B}}T/\sigma$, being of the same magnitude of the driving force \cite{SQUIRES2005}. Thereby, uniform values of the effective friction coefficients are expected at large $Pe$. 

According to Ref.~\cite{SQUIRES2005}, at low $Pe$ the microviscosity increases with the volume fraction, that is $\Delta \eta_{\text{eff}}|_{Pe\rightarrow 0}=2\phi$, and  the effective friction can be written as 

\begin{equation}
\label{eq34}
    \frac{\gamma_{\text{eff}}}{\gamma_0}\bigg|_{Pe\rightarrow 0}=1+2\phi.
\end{equation}

\begin{figure}[h!]
	\centering
		\includegraphics[scale=0.95]{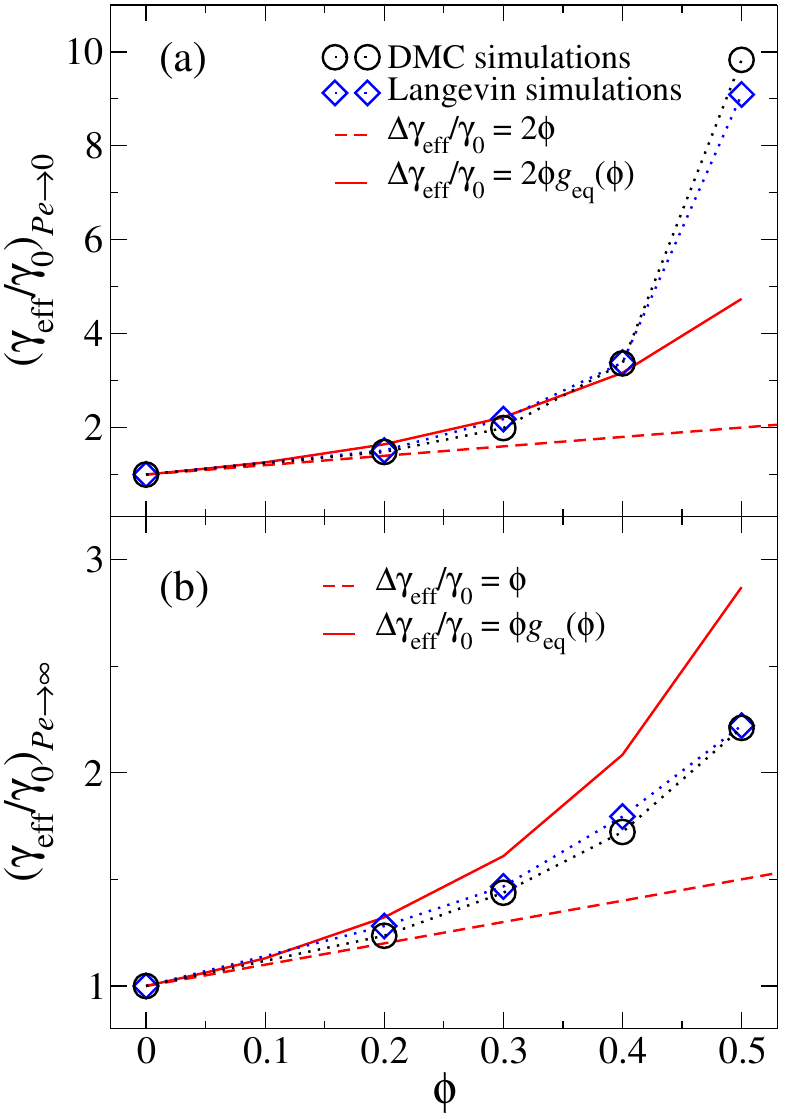}
	\caption{(colour on-line) Effective friction coefficient at (a) low and (b) large $Pe$ limits. Empty circles and diamonds indicate results from DMC and Langevin simulations \cite{Puertas_2014}, respectively. Effective friction coefficients from DMC simulations at the low-$Pe$ and high-$Pe$ limits for each volume fraction were estimated by averaging $\gamma_{\text{eff}}/\gamma_0$ for $Pe\leq5$ and $Pe\geq200$, respectively. Dashed lines represent the theoretical prediction by \citeauthor{SQUIRES2005} \cite{SQUIRES2005}. Solid lines correct these predictions with the equilibrium radial distribution function obtained from independent MC simulations. Dotted lines are guides for the eye.}
	\label{FIG:4}
\end{figure}

\noindent Upon averaging the effective friction coefficients given in Fig.\,\ref{FIG:3} at low $Pe$, in the top frame of Fig.\,\ref{FIG:4} we compare the tendencies observed in our DMC simulations with the theoretical prediction of Eq.~\ref{eq34} and with Langevin dynamics simulations \cite{Puertas_2014}. We observe an excellent quantitative agreement between DMC and Langevin simulations. By contrast, both sets of simulations and theoretical predictions only agree at relatively low volume fractions. To correct the discrepancy at high bath concentrations, \citeauthor{SQUIRES2005} \cite{SQUIRES2005} proposed to rescale the microviscosity with the equilibrium bath particle radial distribution function at the particle-particle contact point, $g_{\text{eq}}(1;\phi)$. This correction has been suggested to counter-balance the reduction of the long-range diffusion coefficient at high volume fractions, whose approximate form is given by $D_s^\infty/D_s\sim (\phi g_{\text{eq}}(1;\phi))^{-1}$. Here we have evaluated $g_{\text{eq}}(1;\phi)$ from independent DMC simulations of spherical particles interacting via the quasi-hard potential of Eq.\,\ref{eq20} (see Fig.~S1 of the Supplementary Material for details). Incorporating this correction significantly improves the agreement between simulations and theory up to $\phi=0.40$, but it is still insufficient to bridge the gap at larger volume fractions. 
At large $Pe$, the microviscosity increments change more softly with the volume fraction and are given by $\Delta \eta_{\text{eff}}|_{Pe\rightarrow \infty}=\phi$. Therefore, the effective friction reads \cite{SQUIRES2005} 

\begin{equation}
\label{eq35}
    \frac{\gamma_{\text{eff}}}{\gamma_0}\bigg|_{Pe\rightarrow \infty}=1+\phi.
\end{equation}

\noindent We notice that our DMC simulations are in excellent agreement with Langevin dynamics simulations also at high $Pe$ as shown in the bottom frame of Fig.\,\ref{FIG:4}. Less satisfactory is the agreement with Eq.\,\ref{eq35}, even when the correction $g_{\text{eq}}(1;\phi)$ is introduced.

\begin{figure}
	\centering
		\includegraphics[scale=0.6]{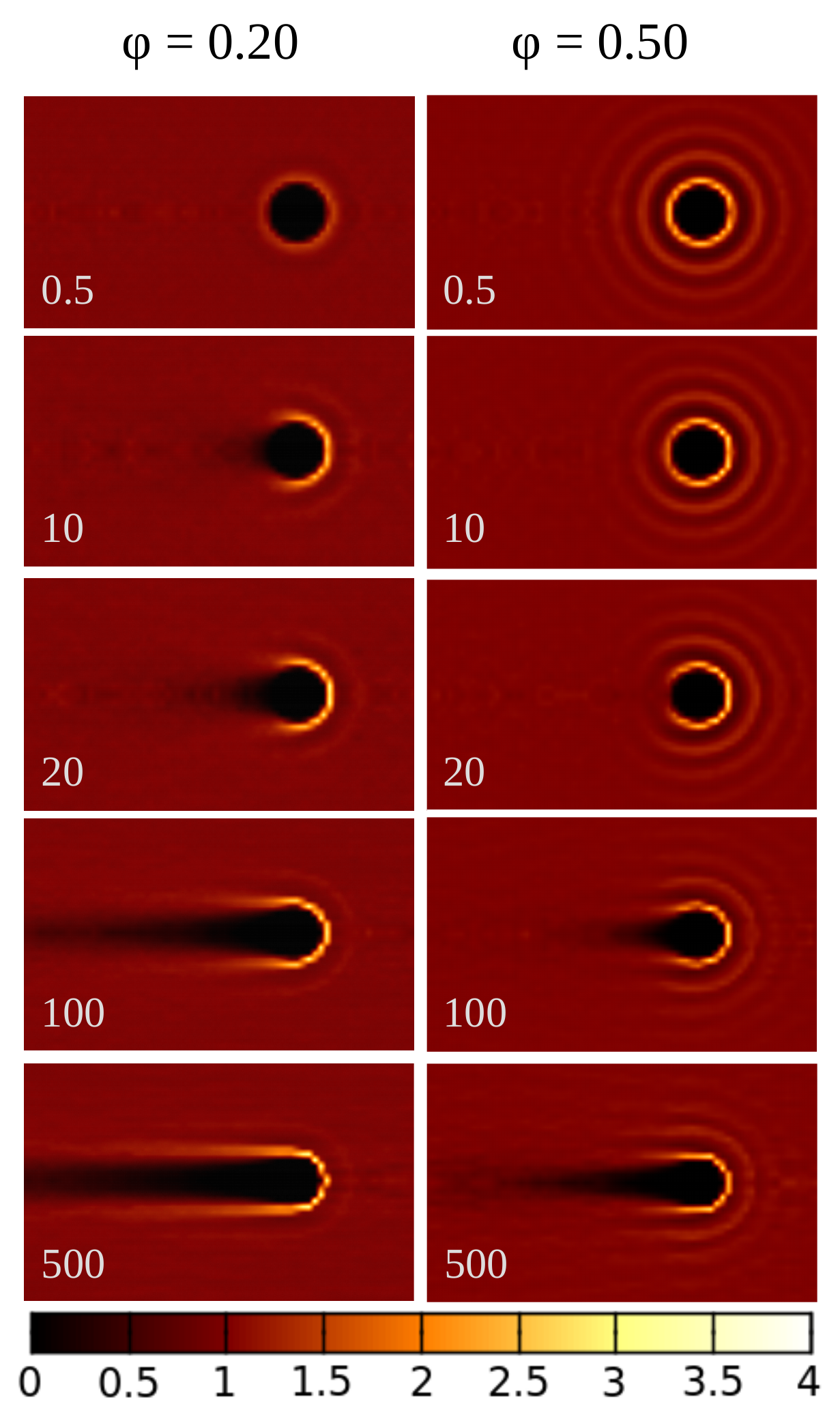}
	\caption{(Colour on-line) Density maps of a colloidal suspension of quasi-hard spheres around a probe particle pulled by a constant force at $\phi=0.20$ (left column) or $\phi=0.50$ (right column) and different P\'eclet numbers as reported in each frame. The colour palette is shown at the bottom of the figure.}
	\label{FIG:5}
\end{figure}

In Fig.~\ref{FIG:5}, we provide the density maps of the host fluid in the vicinity of the tracer particle at different $Pe$ numbers and $\phi=0.20$ (left panels) or $\phi=0.50$ (right panels). At low $Pe$, thermal motion dominates over advection and the structural organisation of the bath particle is basically unperturbed by the motion of the tracer. In this regime, where the asymmetry induced by the driving direction is negligible, the bath density resembles that observed in the absence of external forces (undisturbed Brownian motion). 
As $Pe$ increases, advection progressively dominates over Brownian motion, and force thinning causes a break in the symmetry of the bath density by forming a denser layer of bath particles at the front of the tracer, whilst a low-density wake depleted of bath particles forms behind it. The length of this trailing wake increases with $Pe$, being similar to that observed in past theoretical works \cite{SQUIRES2005, SWAN2013}, simulations \cite{CARPEN2005}, and constant-velocity MR experiments on a bath of polymethyl methacrylate particles \cite{MEYER2006, SRIRAM2010}. Equally interesting is how the bath density influences the response of the local microstructure on the tracer motion. At $\phi=0.20$, the low-density wake behind the tracer persists over a few particle diameters, especially so at $Pe=500$, where its extension approximately achieves $4\sigma$ in the direction of the force applied. This behaviour is less significant at $\phi=0.50$, where the temporary concentration gradient forming behind the moving tracer is too large to persist over longer length scales.

\subsection{Bath of colloidal hard rods}

\begin{figure}[h!]
	\centering
		\includegraphics[scale=0.95]{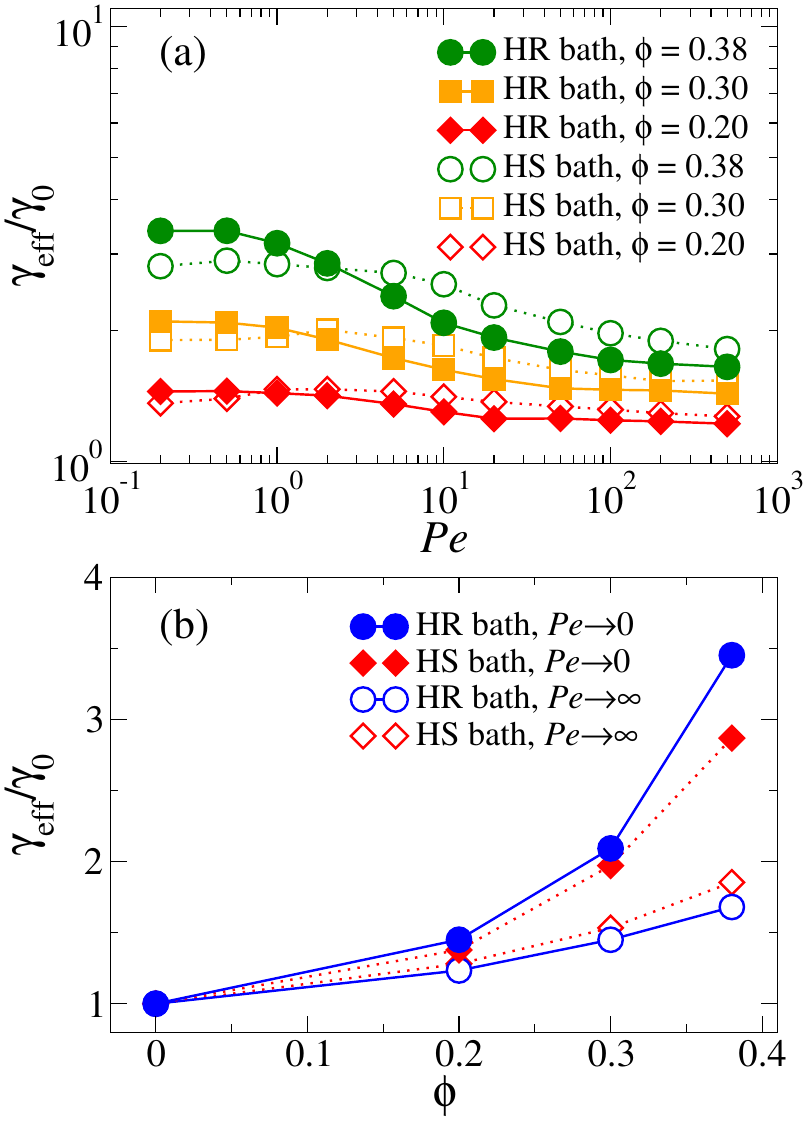}
	\caption{(Colour on-line) (a) Effective friction coefficient as a function of the external force and volume fractions. Solid and empty symbols refer to HR and HS bath, respectively. (b) Effective friction coefficients for HR (circles) and HS (diamonds) baths in the low (solid symbols) and high (empty symbols) $Pe$ limits. The solid and dotted lines are guides for the eye.}
	\label{FIG:6}
\end{figure}

In the previous section, we have shown that DMC can be applied to study constant-force MR in systems of spherical particles. Now we turn our attention to colloidal suspensions of anisotropic particles and consider the case of a spherical tracer driven by an external force through a bath of hard rods (HRs), which are modelled as spherocylinders of aspect ratio $L^*=5$. In this case, all inter-particle interactions are defined via a hard-core potential (Eq.\,\ref{eq21}). While the P\'eclet number still lies in the range $0.2\leq Pe\leq 500$, the volume fractions, $\phi=0.2$, 0.3, and 0.38, have been set to reproduce isotropic phases below the isotropic-to-nematic phase transition \cite{BOLHUIS1997}. Table S2 of the Supplementary Material presents the details of the systems studied. The top frame of Fig.\,\ref{FIG:6} shows the effective friction coefficient calculated by DMC simulations in a bath of HRs as a function of $Pe$. For comparison, we have also included the DMC results corresponding to a bath of hard spheres (HS), which behave similarly to the quasi-hard spheres studied in Section 4.1. As a general tendency, we observe a very similar qualitative behaviour of the effective friction coefficient with varying volume fraction in both HR and HS baths, and this behaviour closely reminds that reported in Fig.\,\ref{FIG:3} for quasi-hard spheres. In particular, we observe a plateau at low and large $Pe$ with a force-thinning regime developing at intermediate forces.

We notice that at low $Pe$ the friction coefficient of the HR bath is slightly larger than that of the HS bath, especially so at $\phi=0.38$. This is evidenced by the solid symbols in the bottom frame of Fig.~\ref{FIG:6}. Since in this regime $\gamma_{\text{eff}}$ is expected to have a very similar value to that in the absence of external forces, then $(\gamma_{\text{eff}})^{-1}\simeq(\gamma_{Pe=0})^{-1}\sim D_s^{\infty}$, where $D_s^{\infty}$ is the long-time self-diffusion coefficient. By performing equilibrium DMC simulations of a spherical tracer in HR and HS baths, we found that the ratio of the long-time tracer's diffusion coefficient at $\phi=0.20$ and 0.38 is $D^{\infty}_{\text{HS}}/D^{\infty}_{\text{HR}}=1.01$, and 1.04, respectively. This indicates that $\gamma_{Pe=0}^{\text{HR}}>\gamma_{Pe=0}^{\text{HS}}$ as indeed observed in the upper and bottom frame of Fig.~\ref{FIG:6} at low P\'eclet numbers. At larger $Pe$, between 1 and 5 depending on $\phi$, we observe a crossover, with the friction coefficients of the two species inverting the tendency observed at low $Pe$. In particular, $\gamma_{\text{eff}}^{\text{HR}}$ becomes lower than $\gamma_{\text{eff}}^{\text{HS}}$ and this behaviour is maintained up to the linear regime at large $Pe$. We believe that this inversion is due to a change in the distribution of the bath particles immediately in front of the tracer. The barrier formed by these particles in both HR and HS baths hampers the tracer's diffusion in the direction of the applied force. By pushing them out of the way, the tracer modifies their local organisation in such a manner that, especially at large volume fractions, the resistance offered to its flow by HSs becomes larger than that offered by HRs. Because the packing of the bath particles is strongly determined by their shape anisotropy, it is reasonable to expect that different particle anisotropies will provide different friction coefficients. To support these arguments, we performed DMC simulations (not shown here) of a spherical tracer diffusing in a bath of shorter HRs at $Pe=20$ and $\phi=0.38$. More specifically, for HRs with $L^*=2.5$, we found $\gamma_{\text{eff}}/\gamma_0=2.0$, a value that is in between that obtained, under the same conditions, for larger rods ($L^*=5$, $\gamma_{\text{eff}}/\gamma_0=1.9$) and spheres ($L^*=0$, $\gamma_{\text{eff}}/\gamma_0=2.3$). This trend in the effective friction coefficient is also observed in the long $Pe$ limit as indicated by the empty symbols in the bottom frame of Fig.~\ref{FIG:6}.

According to the analysis by \citeauthor{SQUIRES2005} \cite{SQUIRES2005} the microviscosity increment for fixed force reads $\Delta\eta_{\text{eff}}\sim (Pe)^{-1}\int n_x g_{\text{M}}(1;\phi) d\Omega$, where $n_x$ is the parallel component to the external force of the normal to the tracer in the outward direction, and $g_{\text{M}}(1;\phi)$ refers to the tracer-bath (M = HR or HS) pair correlation function at contact. Our simulations revealed that upon increasing $Pe$, $g_{\text{HR}}(1;\phi)$ is progressively lower than $g_{\text{HS}}(1;\phi)$. For instance, at $\phi=0.38$ and $Pe=500$, our calculations yield $g_{\text{HS}}(1;\phi)=2.7$ while $g_{\text{HR}}(1;\phi)=1.6$ at the front of the tracer. Thereby, the probability density of a spherical tracer to be in contact with a bath particle is larger in a host medium of HS than in one of HR. Consequently, $\Delta\eta_{\text{eff}}$ will be larger for the HS bath than for the HR bath, which implies that $\gamma_{\text{eff}}^{\text{HS}}>\gamma_{\text{eff}}^{\text{HR}}$.

\begin{figure}[h!]
	\centering
		\includegraphics[scale=0.6]{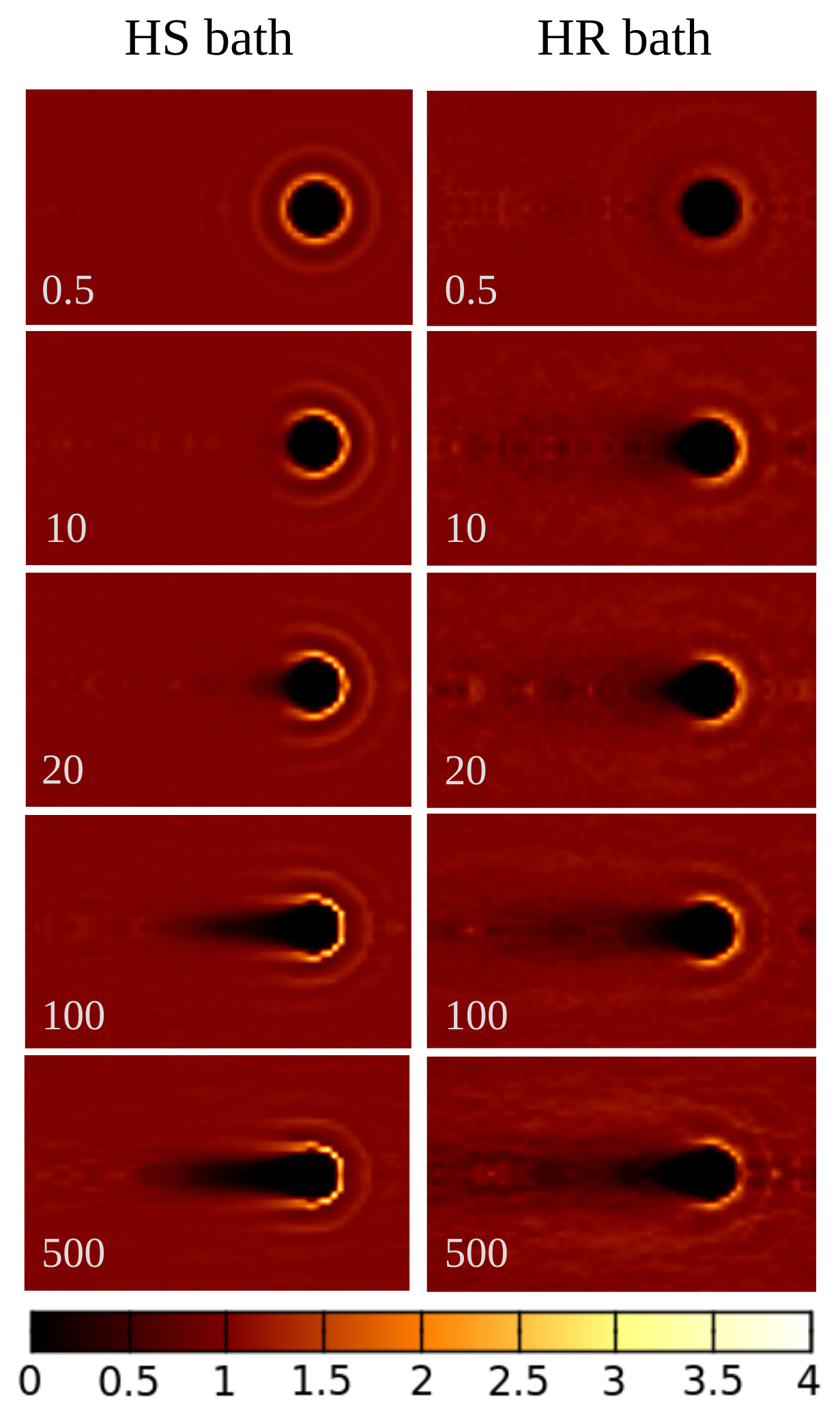}
	\caption{(Colour on-line) Density maps of a colloidal suspension of hard spheres (left) and hard rods (right) around a probe particle pulled by a constant force at $\phi=0.38$ and different Péclet numbers as reported in each frame. The colour palette is shown at the bottom of the figure.}
	\label{FIG:7}
\end{figure}

The average HS and HR bath densities around the tracer particle for $\phi=0.38$ and different P\'eclet numbers are presented in Fig.~\ref{FIG:7}. At very low $Pe$, Brownian motion dominates and counteracts the effects of the tracer, and the host density around the tracer is basically equivalent to the equilibrium bath density. As $Pe$ increases, the force thinning of the effective friction promotes an increased asymmetry of the bath particles distribution around the tracer. While the concentration of the bath particles in front of the tracer increases, a wake depleted of bath particles is formed in the backside of the tracer. As observed for quasi-hard spheres (Fig.~\ref{FIG:5}), the extent of the wake increases with the intensity of the external force. Such a particle-free zone behind the tracer appears slightly larger in the HS bath than in the HR bath, suggesting a more significant change in the local microstructure of the HS bath as compared to that of the HR bath. Interesting differences are also detected in front of the tracer, where $g_{\text{HR}}(1;\phi)<g_{\text{HS}}(1;\phi)$, suggesting a lower probability of tracer-HR collisions as compared to the tracer-HS collisions and thus resulting in a lower effective friction coefficient.

In addition to calculating the local changes in the host phase density upon tracer transit, we have estimated the change in the orientation of the bath particles in regions close to the tracer. Since the particles in the HR bath are in isotropic phase and no preference in their orientation is observed, the emergence of ordering for rods in volumes $\mathbf{r}$ near the tracer can be determined with orientational correlation function, $E_2(\mathbf{r})$. To define the volumes $\mathbf{r}$, the space has been divided into a set of virtual rings centred along the axis of the tracer. This choice is required by the fact that the external force induces an axial symmetry with the movement of the tracer (see Fig.~\ref{FIG:7}). In particular, we have evaluated $E_2(\mathbf{r})$ as the dot product between the unit orientation vectors of $N_{\mathbf{r}}(t)$ particles, $\hat{\mathbf{u}}_i(t)$, within the volume $\mathbf{r}$ at time $t$ and the unit orientation vector, $\hat{\mathbf{F}}_{\text{ext}}$, of the external force:    

\begin{equation}
\label{eq36}
    E_2(\mathbf{r}) = \left<\frac{1}{N_{\mathbf{r}}(t)}\sum_{i=1}^{N_{\mathbf{r}}(t)}\frac{1}{2}\left\{3[\hat{\mathbf{u}}_i(t)\cdot \hat{\mathbf{F}}_{\text{ext}}]^2-1\right\}\right>    
\end{equation}
    
\noindent where $\left<...\right>$ indicates time average. If the rods are locally aligned in the direction of the external force then $E_2(\mathbf{r})\simeq 1$. By contrast, if the particle orientation is perpendicular to $\hat{\textbf{F}}_\text{ext}$ then $E_2(\mathbf{r})\simeq -1/2$. The variation of $E_2({\textbf{r}})$ with $Pe$ at $\phi=0.38$ is depicted in Fig.~S2 of the Supplementary Material. Interestingly, there is a weak ordering over the surface of contact between tracer and bath particles. However, this arrangement corresponds to a geometrical effect and is not a consequence of the motion of the tracer. While the HRs immediately at the front or at the back of the tracer are inevitably perpendicular to the force ($E_2({\textbf{r}})\simeq -1/2$), the particles above and below the tracer can also be aligned with the force ($-1/2\leq E_2({\textbf{r}})\leq 1$). Although this effect is observed for all the volume fractions and $Pe$ considered in this work, the ordering of the HRs around the tracer particle tends to vanish at large P\'eclet numbers. On the other hand, our results do not indicate a local ordering of the bath particles beyond the surface of contact of the tracer since $E_2({\bf{r}})\approx 0$. Most likely, the lack of local ordering beyond the tracer's surface is due to the small size of the tracer as compared to the size of the bath particles surrounding it. If the tracer was larger, above a given $Pe$ number, HRs should orient. The effect of the tracer size on the reorientation of the bath particles is currently under investigation.



\section{Conclusions}
\label{sec5}

In summary, we have generalised our DMC methodology to investigate active MR of colloidal suspensions. More specifically, we considered a tracer particle pulled by an external constant force through colloidal suspensions of spherical and rod-like particles at different $Pe$ numbers and volume fractions. To validate our methodology, we have performed a comparative analysis between DMC and Langevin dynamics simulations \cite{GAZUZ2009, Puertas_2014} for the case of a tracer displacing in a bath of quasi-hard spherical particles. To compare these two techniques, we computed the effective friction coefficient ($\gamma_{\text{eff}}$) experienced by the tracer, which is a measure of the linear and nonlinear local viscoelastic behaviour of the system. We have monitored this parameter by tracking the displacement of the tracer, which is used to obtain the average velocity resulting from the applied external force. This is then employed to define the effective microviscosity as $\gamma_{\text{eff}}=F_{\text{ext}}/\langle v\rangle $. For DMC simulations, we have calculated the average velocity as a function of MC cycles and then rescaled the results onto the BD time scale. The agreement between the two simulation techniques is excellent. In particular, DMC simulations were able to capture the dependence of $\gamma_{\text{eff}}$ on the magnitude of the external force: a plateau at small $Pe$, force thinning as intermediate $Pe$, and a second plateau at large $Pe$. These regimes are tightly connected to the distortion of the microstructure around the moving tracer: a nearly symmetric host particle distribution around the tracer at low $Pe$, then followed by the formation of a dense domain in front of the tracer at intermediate $Pe$, and the eventual  development of a trailing wake behind the tracer at large $Pe$. All these features have also been detected by simulations \cite{GAZUZ2009,CARPEN2005,Puertas_2014} and predicted by theory \cite{SQUIRES2005,SWAN2013}. Validating our DMC method with existing results was a crucial preliminary step to apply it to investigate the MR of hard rod-like particles in isotropic phases. Although the qualitative dependence of the effective friction coefficient on the applied force is similar to the case of a bath of hard spheres, we observe quantitative differences that are related to the geometry of the host particles and their ability to pack around the moving tracer. While near equilibrium (low $Pe$) $\gamma_{\text{eff}}^{\text{HS}}<\gamma_{\text{eff}}^{\text{HR}}$, at intermediate and large forces, where the bath is deformed by the tracer, the opposite situation is observed. For instance, at $\phi=0.38$ and $Pe=500$ the pair correlation function of spheres $g_{\text{HS}}(1;\phi)$ at the contact point in front of the tracer is 1.7 times larger than its analogue for a bath of rods $g_{\text{HR}}(1;\phi)$. This would indicate that the occurrence of a tracer-HS contact is more likely than a tracer-HR contact, resulting in $\gamma_{\text{eff}}^{\text{HS}}>\gamma_{\text{eff}}^{\text{HR}}$.  

The DMC simulation results presented in this work show an excellent agreement with the Langevin dynamics simulation results in the study of constant-force MR of colloidal systems. Nevertheless, we would like to remind that DMC has been applied under some important assumptions. At low P\'eclet numbers, the DMC method allows the use of MC time steps in the range $\sim10^{-3}\tau-10^{-1}\tau$ that, after rescaling (see Eqs.~\ref{eq7} or \ref{eq18}), are significantly larger than those regularly employed in BD and Langevin dynamics simulations. This opens up the possibility of exploring longer time scales, which are especially relevant in particularly dense systems, at small to intermediate force intensities \cite{ORTS20198, ORTS2020052607,PUERTAS2018}. By contrast, in the large $Pe$ limit, only small time steps are allowed ($\sim10^{-7}\tau-10^{-6}\tau$). This constraint emerges from the condition $\beta F_{\text{ext}}\delta x\ll 1$ (Eq.~\ref{eq9}). Since the maximum displacement of the tracer can be approximated by $\delta x\simeq\sqrt{2D_t\delta t_{\text{MC},t}}$, the tracer time step should be small enough to satisfy the aforementioned condition. Therefore, in order to increase the range of time steps at large forces, a larger number of terms should be considered in the expansion of the averaged quantities in section \ref{sec22}. Finally, the DMC method ignores the solvent mediated hydrodynamic interactions, which might become important in the limit of large forces \cite{SWAN2013,Nazockdast2016733}. Taking into account these considerations, the extended DMC presented in this work, represents an effective methodology to qualitatively and quantitatively study the constant-force MR behaviour of colloidal suspensions.


\section*{Acknowledgements}
F.A.G.D., A.M.P and A.P. acknowledge the International Exchanges Grant IES\textbackslash R1\textbackslash 191066, awarded by The Royal Society. F.A.G.D. and A.P. also acknowledge the Leverhulme Trust Research Project Grant RPG-2018-415, and the assistance given by Research IT and the use of the Computational Shared Facility at The University of Manchester. A.M.P. was also funded by the Spanish Ministerio de Ciencia, Innovaci\'on y Universidades under project PGC2018-101555-B-I00 and UAL/CECEU/FEDER through project UAL18-FQM-B038-A.

\bibliographystyle{model1-num-names}
\bibliography{cas-refs}







\end{document}